\documentclass[%
 reprint,
 superscriptaddress,
 amsmath,amssymb,
 aps	%submitted to prapplied
]{revtex4-2}

\usepackage[utf8]{inputenc}
\usepackage[T1]{fontenc}
\usepackage{mathptmx}   % change font family

\usepackage{graphicx}% Include figure files
\usepackage{bm}% bold math
\usepackage{mathrsfs}
\usepackage{siunitx}
\usepackage[hidelinks,
    colorlinks=true,
    linkcolor=blue,     % color of internal links 
    citecolor=blue,     % color of links to bibliography
    filecolor=blue,     % color of file links
    urlcolor=blue]{hyperref}% add hypertext capabilities
\usepackage{ORCIDinREVTeX}

\renewcommand{\eqref}[1]{\hyperref[{#1}]{\textup{(\ref*{#1}})}}
\newcommand{\figref}[1]{\hyperref[{#1}]{\textup{Fig.~\ref*{#1}}}}
\newcommand{\appref}[1]{\hyperref[{#1}]{\textup{Appendix~\ref*{#1}}}}
\newcommand{\tilkappa}{\tilde{\kappa}}
\allowdisplaybreaks

\begin{document}
\title{Ultrahigh-Resolution Wireless Capacitance Readout Based on Single Real Mode \\ 
    in Perturbed Parity-Time-Symmetric Sandwich-Type Electronic Trimer}

\author{Ke Yin}
\orcid{0000-0002-8534-216X}

\author{Yuangen Huang}
\orcid{0000-0002-2486-2778}

\author{Wenjing Yin}
\orcid{0000-0001-9412-7114}

\author{Xianglin Hao}
\orcid{0000-0001-7149-0113}

\author{Xikui Ma}

\author{Tianyu Dong}
\orcid{0000-0003-4816-0073}
\email[Author to whom correspondence should be addressed. Please e-mail to: ]{\mbox{tydong@mail.xjtu.edu.cn}}
\affiliation{School of Electrical Engineering, Xi'an Jiaotong University, Xi'an 710049, China.}

\date{\today}

\begin{abstract}
High-performance interrogation of inductor-capacitor (LC) microsensor has been a long-standing challenge due to the limited size of the sensor. Parity-time (PT) symmetry, an intriguing concept originated from quantum physics, has been utilized to improve the spectral resolution and sensitivity of the conventional readout circuit, while the PT symmetry condition has to be satisfied. In this work, a sandwich-type wireless capacitance readout mechanism based on perturbed PT-symmetric electronic trimer without manual tuning of the reader circuit is proposed. Theoretical eigenvalue analysis by solving the system equations shows that the system exhibits single real mode in weak coupling regime whose eigenfrequency changes in response to the capacitance of the neutral resonator. Furthermore, the proposed readout system exhibits wider readout capacitance range compared to standard PT-symmetry-based system while retaining higher $Q$-factor compared to conventional readout method, which has been validated with the experimental prototype based on printed circuit board. Our work not only enriches the underlying theory of non-Hermitian physics, but also shows potential applications in scenarios where longer interrogation distance is required, such as implanted medical devices, parameter detection in harsh environment, \emph{etc}.
\end{abstract}

\maketitle

\section{\label{sec:intro}Introduction}
Thanks to the capability of remote inquiry and the unnecessity of additional power supply, inductor-capacitor (LC) passive sensor has been utilized for implanted medical devices to monitor physiological parameters of patients such as intraocular pressure (IOP) \cite{chen2010wireless}, blood flow \cite{boutry2019biodegradable}, wound pressure \cite{deng2018lc}, and for parameter detection in harsh environment such as pressure \cite{rogers2017passive}, temperature \cite{ji2018novel} and liquid permittivity \cite{zarifi2017high}, \emph{etc}. Conventional LC wireless sensing system typically consists of a readout antenna and an inductively coupled passive RLC microsensor whose resonance frequency change, in response to the quantities to be measured, is wirelessly interrogated through the dip frequency shift of the reflection spectrum \cite{nopper2009wireless}. However, due to the large equivalent resistance caused by skin effect and eddy current loss, the real-life application of LC microsensor has been limited by the low quality ($Q$)-factor and poor resolution of the readout reflection spectrum. Moreover, the small size of microsensor results in weak coupling between the readout and the sensor coils, which further degrades the readout performance. In order to improve the readability and robustness of detection, the measured reflection dip should generally be sharp and narrow band.

Aiming to the above issue, parity-time (PT) symmetry has provided a solution to greatly enhance the $Q$-factor of the reflection spectrum \cite{chen2018generalized}. The concept of PT symmetry is first presented in the context of quantum physics to describe a new class of non-Hermitian Hamiltonians with real energy spectrum \cite{bender1998real}. In recent years, such an appealing discovery has been fruitfully extended to other research areas including photonics \cite{guo2009observation,ruter2010observation,ozdemir2019parity}, electronics \cite{schindler2011experimental,schindler2012symmetric,assawaworrarit2017robust,zhou2018nonlinear,assawaworrarit2020robust,chen2018generalized,dong2019sensitive,sakhdari2019experimental,xiao2019enhanced,yang2021ultrarobust,yang2022vitro,yin2022wireless}, microwaves \cite{bittner2012p,fan2020hybrid}, and acoustics \cite{fleury2015invisible,fleury2016parity}. Among them, non-Hermitian electronics has become an active research area, and opens up interesting avenues for applications such as robust wireless power transmission (WPT) \cite{assawaworrarit2017robust,zhou2018nonlinear,assawaworrarit2020robust,hao2022enhanced,hao2022frequency} and wireless microsensor readout with enhanced sensitivity and resolution \cite{chen2018generalized,sakhdari2019experimental,yang2021ultrarobust,yang2022vitro}. 

This non-Hermitian PT-symmetry-enhanced wireless sensing system can typically work in two phases, \emph{viz.}, PT-symmetric phase and PT-broken phase. Standard second-order PT symmetric telemetry system consists of two inductively coupled RLC resonators with balanced gain and loss, which has been utilized for interrogating resistive temperature sensors \cite{yang2021ultrarobust,kananian2020coupling} and intracranial pressure sensors \cite{yang2022vitro}, \emph{etc}. The second-order configuration can further be extended to third-order system by incorporating three inductively coupled gain-neutral-loss LC resonators chain, in which a divergent exceptional point (DEP), \emph{i.e.}, $\kappa_\text{D} = 1/\sqrt{2}$, enables an infinity sensitivity in theory \cite{sakhdari2019experimental}. Therefore, by deliberately manipulating the system to work nearby DEP, \emph{i.e.}, $\kappa \approx 0.7$, the system can realize ultrasensitive wireless sensing. Nevertheless, such strong coupling has approached the upper limited of three linearly coupled inductors [see \appref{sec:append_c} for theoretical verification], which is not easy to realize in practical applications. Furthermore, these symmetry-phase-based readout systems require manual tuning of the reader circuit to satisfy the PT-symmetry condition and the coupling coefficient larger than $\kappa_\text{EP}$ to make the system operate in the PT-symmetric phase where high sensitivity and $Q$-factor occurs.

Along different lines, it has been shown that the PT-symmetry breaking regime can enhance the remote interrogation distance while maintaining high $Q$-factor \cite{zhou2020enhancing}. However, cyclically scanning variable capacitance on the reader side utilized to track the change of the capacitive humidity sensor increases the system complexity and is time-consuming. Coupling-independent and real-time wireless resistive sensing system based on nonlinear PT symmetry has been reported to realize low-complexity readers \cite{kananian2020coupling}. Moreover, eigenfrequency evolution in response to the sensor-side capacitive perturbation when the system is biased at the exceptional point is investigated, showing that the eigenfrequency bifurcation follows a square-root-dependence on the capacitive perturbation parameter which can be utilized to detect small signal changes \cite{zhou2021observation}. Therefore, EP-enhanced capacitance readout can be utilized without the need of maintaining PT symmetry. In a later work, our group proposed a vector-network-analyzer-free PT-symmetric capacitance readout system with nonlinear gain in which the reader-side capacitance need not to be manually tuned \cite{yin2022wireless}. However, in both these works \cite{zhou2021observation,yin2022wireless}, the readout sensitivity is not high and has yet to be extended to higher-order PT-symmetric systems which greater sensitivity is envisioned. 

In this work, we report a sandwich-type wireless capacitance readout system with ultrahigh $Q$-factor based on perturbed PT-symmetric electronic trimer. We theoretically analyze the eigenfrequency evolution of the third-order PT-symmetric LC sensing system in response to the capacitive perturbation on neutral resonator, showing that single real mode can exist in the weak coupling regime. Furthermore, the proposed readout mechanism enables much sharper reflection dips and higher $Q$-factor capacitance readout compared to conventional LC sensors, and exhibits wider readout capacitance range compared to the standard second-order PT-symmetric system without the need of manually adjusting the capacitance on the gain/loss side. Our work can enrich the underlying physics of non-Hermitian electronics and pave the way towards applications scenarios where high-performance wireless readout with longer interrogation distance is required.

\section{Theory: Eigenvalue analysis of perturbed PT-symmetric electronic trimer} \label{sec:thoery}
\begin{figure}[!ht]
    \centering
    \includegraphics[width=3.4in]{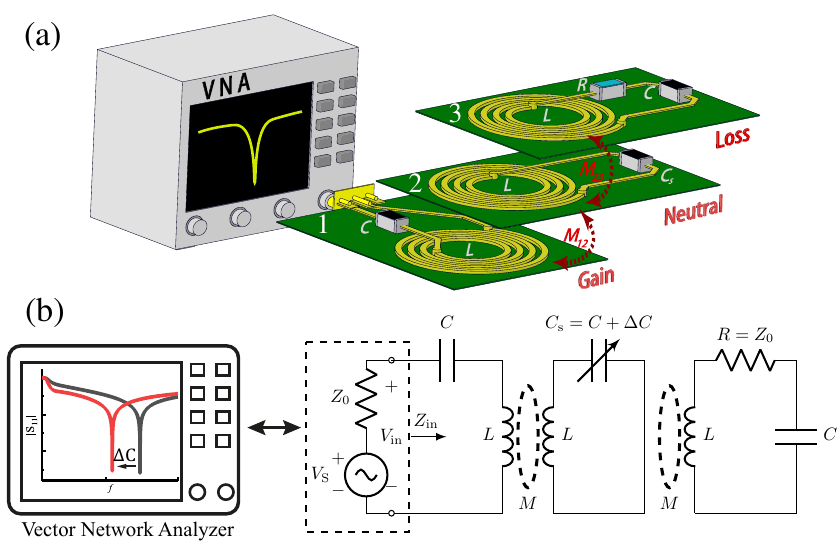}
    \caption{(color online) (a) Illustration of a third-order parity-time symmetric wireless sensing system and (b) the corresponding circuit schematics.}
    \label{fig:schematics}
\end{figure}

Figures \ref{fig:schematics}(a) and \ref{fig:schematics}(b) show the system illustration and the corresponding circuit schematics of the proposed wireless capacitance readout mechanism, respectively. The system is composed of inductively coupled gain-neutral-loss LC resonators chain, which constitutes a PT-symmetric electronic trimer. The gain is introduced by the radio-frequency source of the vector network analyzer (VNA) with characteristic impedance $Z_0 = 50~\si{\ohm}$. The loss is realized with a normal resistor whose resistance is deliberately chosen as $R = Z_0$ to guarantee balanced gain and loss which is typically required in PT-symmetric systems. Different from standard third-order PT-symmetric systems where the capacitance of gain, neutral and loss resonators are tuned together \cite{sakhdari2019experimental}, we consider the case when only the neutral-side capacitance is changed which is regarded as capacitive perturbation to the neutral resonator. Thus, no variation of the gain and loss side is needed to guarantee the parity symmetry as usual.

By defining a state vector $\Phi = (I_1, I_2, I_3, I'_1, I'_2, I'_3)^T$ with $I_n (n = 1, 2, 3)$ denoting the inductor current on the gain, neutral and loss resonators respectively, and $I'_n (n = 1, 2, 3)$ denoting their corresponding first-order derivatives with respect to the normalized time $\tau$ (see \appref{sec:append_a} for the detailed derivation), the system equation can be expressed as
\begin{equation} \label{eq:sys-eq}
  \frac{\text{d} \Phi}{\text{d} \tau} = \mathcal{L} \Phi,
\end{equation}
where the Liouvillian $\mathcal{L}$ reads as
\begin{equation} \label{eq:Liouvillian}
  \mathcal{L} =
  \begin{pmatrix}
    0        & 0          & 0          & 1          & 0        & 0     \\
    0        & 0          & 0          & 0          & 1        & 0     \\   
    0        & 0          & 0          & 0          & 0        & 1    \\
    -\frac{1-\kappa^2}{\tilkappa}  & \frac{\kappa}{\eta \tilkappa}   & -\frac{\kappa^2}{\tilkappa}  & \gamma \frac{1-\kappa^2}{\tilkappa}  & 0 & -\gamma \frac{\kappa^2}{\tilkappa}   \\
    \frac{\kappa}{\tilkappa}   & -\frac{1}{\eta \tilkappa}     &  \frac{\kappa}{\tilkappa}   & -\gamma \frac{\kappa}{\tilkappa}   & 0 & \gamma \frac{\kappa}{\tilkappa}    \\
    -\frac{\kappa^2}{\tilkappa} & \frac{\kappa}{\eta \tilkappa}  & -\frac{1-\kappa^2}{\tilkappa} & \gamma \frac{\kappa^2}{\tilkappa} & 0 & -\gamma \frac{1-\kappa^2}{\tilkappa} \\
  \end{pmatrix}.
\end{equation}  
Here, the parameter $\eta = C_s/C$ reflects the perturbation caused by the capacitive sensor; $\kappa = M/L$ is the mutual coupling coefficient and $\tilkappa = 1 - 2\kappa^2$; $\tau = \omega_r t = t/\sqrt{LC}$ is the normalized time; $\gamma = Z_0 \sqrt{C/L} = R \sqrt{C/L}$ is the normalized gain/loss parameter. It is noted that the corresponding effective Hamiltonian $\mathcal{H}_\text{eff} = \text{i} \mathcal{L}$ is PT-symmetric when $\eta = 1$. That is to say $\mathcal{H}_\text{eff}$ and $\mathcal{PT}$ are commutative, \emph{i.e.}, $[\mathcal{H}_\text{eff},\mathcal{PT}]=0$, where the $\mathcal{P}$ and $\mathcal{T}$ operators respectively reads as
\begin{subequations} \label{eq:pt_operator}
\begin{align} 
    \mathcal{P} &= \begin{pmatrix}
      \bm{J}_3 & 0 \\
      0 & \bm{J}_3 \\
    \end{pmatrix}, \\
    \mathcal{T} &= \begin{pmatrix}
      \bm{I}_3 & 0 \\
      0 & \bm{I}_3 \\
    \end{pmatrix}\mathscr{K},
\end{align}
\end{subequations}
with $\bm{J}_n$ being an $n \times n$ exchange matrix ($J_{ij} = 1$ when $i+j = n+1$ and $J_{ij} = 0$ when $i+j \neq n+1 $), $\bm{I}_n$ being an $n \times n$ identity matrix ($I_{ij} = 1$ when $i=j$ and $I_{ij} = 0$ when $i \neq j$) and $\mathscr{K}$ standing for the complex conjugate. Now, solving the characteristic equation 
\begin{equation} \label{eq:characteristic-eq}
    \text{Det}(\mathcal{L} - \text{i}\omega \bm{I}_6) = 0
\end{equation}
which yields $\omega^6 + c_1 \omega^4 + c_2 \omega^2 + c_3 = 0$ with $c_1 = c_3 [1+\eta(2-\gamma^2-2\kappa^2)]$, $c_2 = -c_3 (2-\gamma^2+\eta)$ and $c_3 = -1/[\eta (1-2\kappa^2)]$, one can obtain the corresponding eigenfrequencies. Note that the characteristic equation is a cubic equation with respect to $\omega^2$, thus the six eigenfrequency solutions are the opposite of each other. In what follows, only positive eigenfrequencies are considered for discussion.

\begin{figure}[!ht]
    \centering
    \includegraphics[width=3.4in]{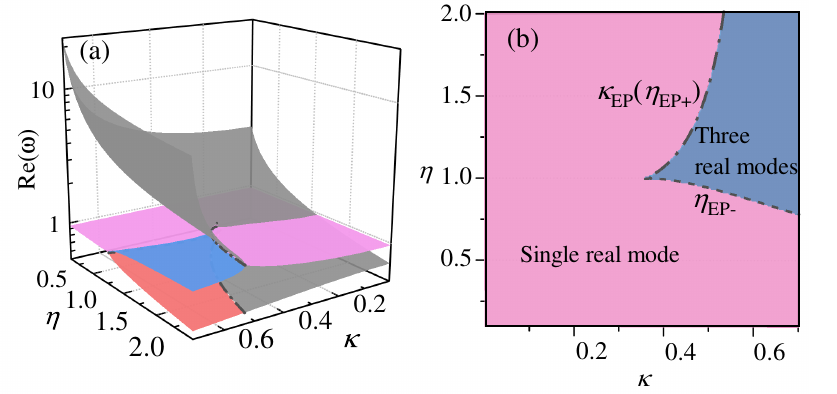}
    \caption{(color online) (a) Real part of system eigenfrequency evolution of perturbed PT-symmetric electronic trimer. (b) The parameter regions for different numbers of real modes in the parameter space $(\kappa,\eta)$.}
    \label{fig:eigenfrequency_evolution_3d}
\end{figure}

For the perturbed PT-symmetric electronic trimer when $\eta \neq 1$, the solution to the characteristic equation \eqref{eq:characteristic-eq} yields the following set of eigenfrequencies \cite{yin2022wireless}, \emph{i.e.},
\begin{subequations} \label{eq:omegas}
  \begin{align}
    \omega_{1,4} & = \pm \left( s + t - c_1/3 \right)^{1/2}, \\
    \omega_{2,5} & = \pm \left[ -(s+t)/2 + \text{i} \sqrt{3}(s-t)/2 - c_1/3 \right]^{1/2}, \\
    \omega_{3,6} & = \pm \left[ -(s+t)/2 - \text{i} \sqrt{3}(s-t)/2 - c_1/3 \right]^{1/2},
  \end{align}
\end{subequations}
where $s = [p+(p^2+q^3)^{1/2}]^{1/3}$ and $t = [p-(p^2+q^3)^{1/2}]^{1/3}$ with $p = -c_1^3/27 + c_1 c_2/6 - c_3/2$ and $q = -c_1^2/9 + c_2/3$. Figs. \ref{fig:eigenfrequency_evolution_3d}(a) and \ref{fig:eigenfrequency_evolution_3d}(b) show the eigenfrequency evolution and the parameter regions for different numbers of real modes at $\gamma = 0.527$ ($C = 100~\si{pF}$) in the parameter space $(\kappa,\eta)$. It is shown that at different $(\kappa,\eta)$, the system can have different numbers of real mode which is determined by the discriminant $\Delta = p^2+q^3$. The system exhibits three real modes when $\Delta < 0$ and single real modes when $\Delta > 0$, which are indicated by the blue region and pink region in \figref{fig:eigenfrequency_evolution_3d}(b) respectively.

\begin{figure}[!ht]
    \centering
    \includegraphics[width=3.4in]{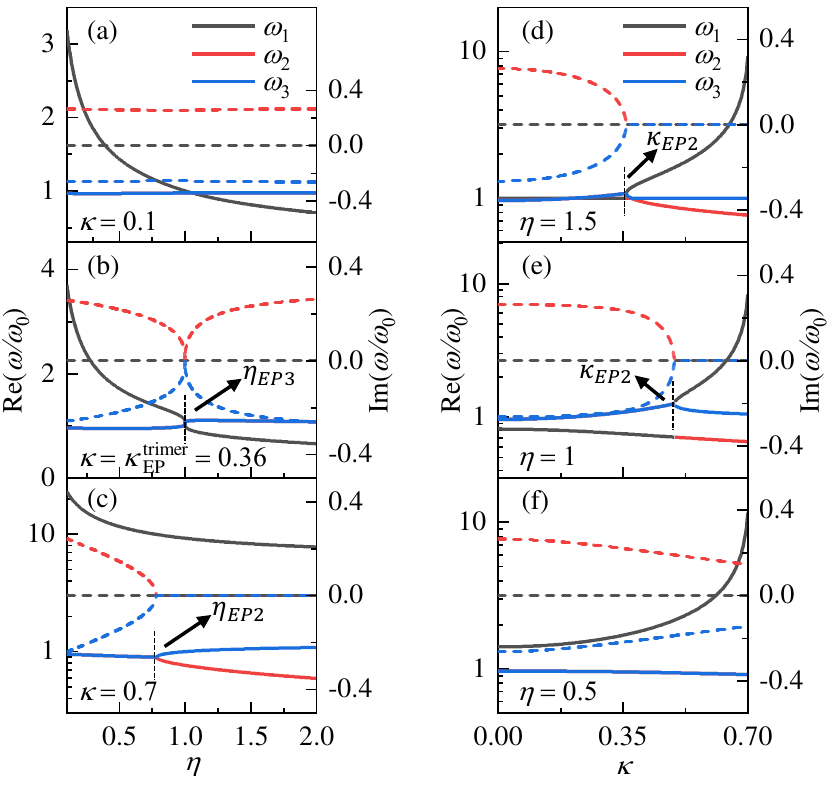}
    \caption{(color online) Real (solid curves) and imaginary (dashed curves) part of eigenfrequency evolution as a function of the perturbation parameter $\eta$ at (a) $\kappa = 0.1$, (b) $\kappa = 0.36$, and (c) $\kappa = 0.7$ respectively. Real and imaginary part of eigenfrequency evolution as a function of the coupling coefficient $\kappa$ at (d) $\eta = 1.5$, (e) $\eta = 1$, and (f) $\eta = 0.5$ respectively.}
    \label{fig:eigenfrequency_evolution_perturbed}
\end{figure}

Figures \ref{fig:eigenfrequency_evolution_perturbed}(a) -- \ref{fig:eigenfrequency_evolution_perturbed}(c) plot the real and imaginary part of the eigenfrequency evolution as a function of the perturbation parameter $\eta$ when the coupling coefficient $\kappa$ reads 0.1, 0.36 and 0.7, respectively. In the strong coupling regime when $\kappa = 0.7$, the eigenfrequency evolution with respect to $\eta$ exhibits two regions: When $\eta < \eta_\text{EP2}$, $\Delta > 0$, the system has one real eigenvalue $\omega_1$ and a pair of conjugate eigenvalues $\omega_{2,3}$; when $\eta > \eta_\text{EP2}$, $\Delta < 0$, the system exhibits three real eigenvalues. The second-order exceptional point $\eta_\text{EP2} = 0.782$ is where $\omega_2$ and $\omega_3$ coalesce and become a 2nd-order degeneration. It should be noted that the real eigenfrequency $\omega_1$ exhibits a dramatic change in response to $\eta$ when $\kappa$ is close to $\sqrt{2}/2$, which is the upper limit of three linearly coupled inductors. When $\kappa = \kappa_\text{EP}^\text{trimer} = \gamma \sqrt{4 -\gamma^2}/(2 \sqrt{2}) = 0.36$, there are only three real eigenfrequencies at the same time when $\eta = \eta_\text{EP3} = 1$ where three modes coalesce and become a third-order degeneration. In addition, the system exhibits only one real mode denoted by $\omega_1$ and a pair of complex conjugate modes denoted by $\omega_{2,3}$. In the weak coupling regime when $\kappa = 0.1$, the system only has one real mode denoted by $\omega_1$ for any given $\eta$; and $\omega_{2,3}$ denote a pair of complex conjugate eigenmodes. In this work, the single real mode $\omega_1$ in weak coupling regime, \emph{i.e.}, $\kappa < \kappa_\text{EP}^\text{trimer} = 0.36$, is utilized for the proposed wireless sensing scheme.

Figures \ref{fig:eigenfrequency_evolution_perturbed}(d) -- \ref{fig:eigenfrequency_evolution_perturbed}(f) illustrate the real and imaginary part of the eigenfrequency evolution as a function of the coupling coefficient $\kappa$ for various perturbation parameter $\eta$ of 1.5, 1 and 0.5, respectively. When $\eta \geq 1$, the eigenfrequency evolution with respect to the coupling coefficient $\kappa$ exhibits two regions: When $\kappa < \kappa_\text{EP2}$, $\Delta > 0$, the system has one real eigenvalue $\omega_1$ and a pair of conjugate eigenvalues $\omega_{2,3}$; when $\kappa > \kappa_\text{EP2}$, $\Delta < 0$, the system exhibits three real eigenvalues. However, when $\eta < 1$, the system only exhibits single real modes even in the strong coupling regime. Therefore, only when the sensor capacitance is larger than the reader capacitance, there could be three real modes at the same time even though the PT symmetry condition is not satisfied.

It should be pointed out that when the perturbation is caused by the loss-side capacitance, the characteristic equation becomes $\omega^6 + c_4\omega^4 + c_2\omega^2 + c_0 + \text{i}(c_3\omega^3 + c_1\omega) = 0$, where $c_4 = c_0 [1-\kappa^2+\eta(2-\gamma^2-\kappa^2)], c_2 = - c_0 [2+(1-\gamma^2)\eta], c_3  = -c_1 = c_0 \gamma(1-\eta), c_0 = -1/[\eta ( 1- 2 \kappa^2)]$. No explicit eigenfrequency solutions exist and the eigenfrequencies can only be solved numerically which are six different complex numbers. Thus, the loss side may not function as the sensor which will break the balanced gain and loss.

\section{Results and Discussions}\label{sec:result}
\begin{figure}[!ht]
    \centering
    \includegraphics[width=3.0in]{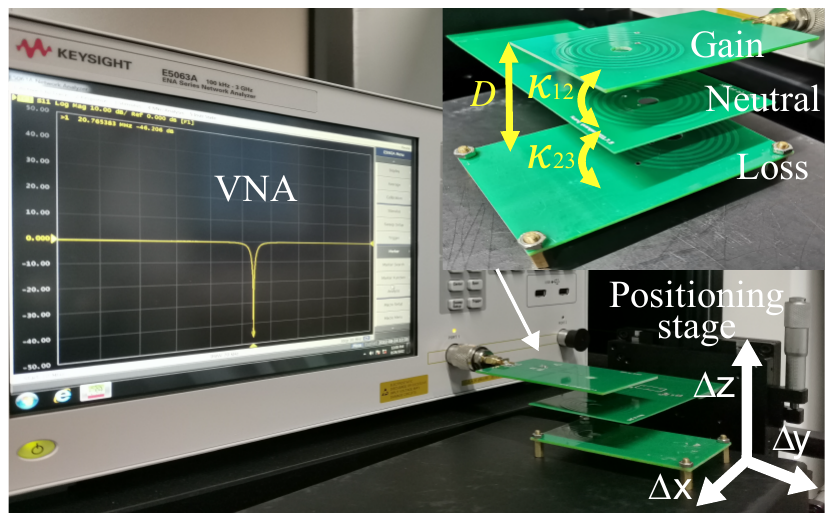}
    \caption{(color online) Experimental setup of the proposed perturbed third-order PT-symmetric wireless sensing system.}
    \label{fig:experiment}
\end{figure}

Figure \ref{fig:experiment} shows the experimental setup of a wireless capacitance readout prototype based on the proposed mechanism. Printed spiral planar inductors with inductance $L=0.9~\si{\micro\henry}$ are manufactured on three printed circuit boards (PCBs), which are coaxially aligned, \emph{i.e.} $\Delta x = \Delta y = 0$; and the adjacent vertical distances $\Delta z$ are adjusted with a positioning stage for coupling parameter tuning (see \appref{sec:append_c} for the detailed information on the coil design). The reflection spectrum $S_\text{11}$ is measured by connecting the gain resonator to a KEYSIGHT E5063A VNA. Neutral LC tank is used as the sensor whose capacitance $C_s$ can be adjusted from $10~\si{pF}$ to $68~\si{pF}$ so as to respond to the capacitance range corresponding to the physical parameters.

\begin{figure}[!ht]
    \centering
    \includegraphics[width=3.4in]{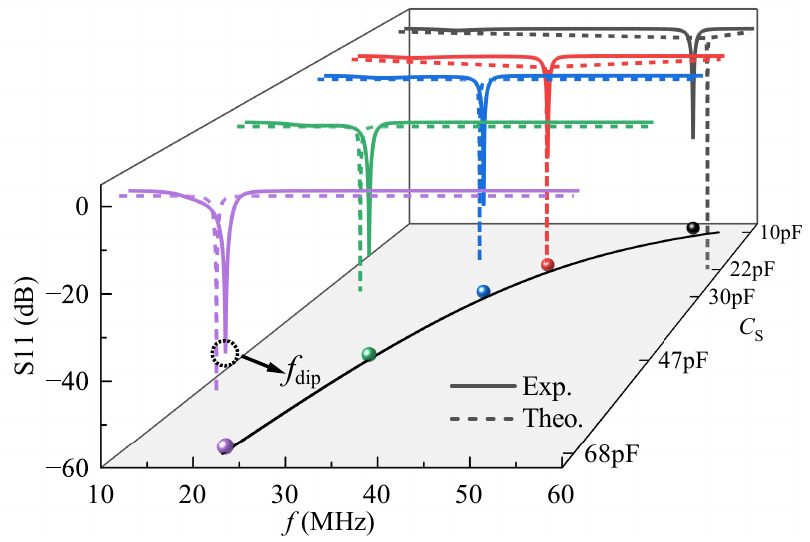}
    \caption{(color online) Theoretical (dashed curves) and experimental (solid curves) of the reflection spectra of the perturbed PT-symmetric electronic trimer when the axial separation distance between the gain and loss coils is $D = 30~\si{mm}$, which yields the coupling coefficient $\kappa = 0.12$. The black solid curve on the $(C_s, f)$ plane presents the theoretical eigenfrequency evolution as a function of sensor capacitance derived from \eqref{eq:omegas}, and markers on the curve correspond to the measured dip frequency projected from the experimental $|S_{11}|$ curves.}
    \label{fig:reflection}
\end{figure}

Figure \ref{fig:reflection} plots the reflection spectra of the proposed wireless capacitance readout mechanism when the coupling coefficients between adjacent resonators are $\kappa_{12} = \kappa_{23} = \kappa = 0.12$. Note that in this coupling configuration, the coupling between the gain and loss coils $\kappa_{13}$ can be neglected. Now, the circuit parameters are: $L = 0.9~\si{\micro\henry}$, $R = Z_0 = 50~\si{\ohm}$ and several discrete capacitance values $C_s$ of 10~\si{pF}, 22~\si{pF}, 30~\si{pF}, 47~\si{pF} and $68~\si{pF}$ are chosen to mimic the capacitive sensor. The solid and dashed curves correspond to the experimental and theoretical results of $S_{11}$, respectively. It is evident that the frequency $f_\text{dip}$ corresponding to the sharp dip with great $Q$-factor is shifting as the sensor capacitance $C_s$ is changed, implying the eigenfrequency of the single real mode changes in response to $C_s$, which validates the aforementioned theoretical analysis. The black solid curve on the $(C_s, f)$ plane illustrates the theoretical eigenfrequency evolution as a function of the sensor capacitance derived from \eqref{eq:omegas}, on which the markers correspond to the experimental dip frequencies of each $|S_{11}|$ spectra. We shall point out that the difference between the experimental and theoretical results will become large when the sensor capacitance is small due to the increase of practical inductance value in high frequency range (see \appref{sec:append_c} for the measured inductance as a function of frequency). Furthermore, the minimum of $|S_{11}|$ can reach $-40~\si{dB}$, which is sufficient enough for capacitance readout. In the experiment, the axial separation distance between the concentric planar spiral inductors on gain and loss side is $D = 30~\si{mm}$, yielding the coupling coefficient $\kappa = 0.12$, which indicates that the proposed readout mechanism can provide a high-resolution readout even in weak coupling regime.

\begin{figure}[!ht]
    \centering
    \includegraphics[width=3.4in]{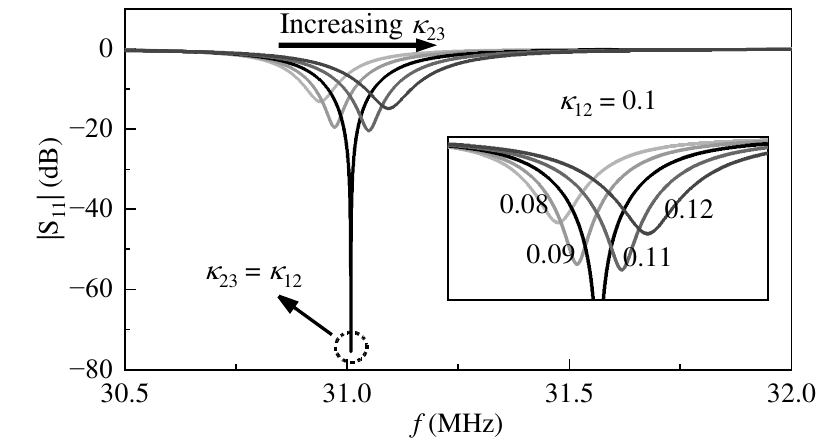}
    \caption{$|S_{11}|$ spectra for the PT-symmetric trimer with asymmetric coupling. When fixing $\kappa_{12}$ to 0.1 and changing $\kappa_{23}$ from 0.08 to 0.12, the sharpest curve occurs when $\kappa_{23}$ is coincident with $\kappa_{12}$.}
    \label{fig:impact-of-kappa}
\end{figure}

For the PT-trimer-enhanced wireless readout system, a crucial part is to guarantee balanced coupling coefficient between adjacent inductor coils. In the case of asymmetric coupling when the coupling coefficients of the gain-neutral and neutral-loss coils are not equal, \emph{i.e.}, $\kappa_{12} \neq \kappa_{23}$, the $Q$-factor will degrade significantly even only a slight unbalance of coupling between adjacent coils exhibits. Theoretical results show that when $\kappa_{12}$ is fixed and as $\kappa_{23}$ is changed, the reflection spectrum curve with the greatest $Q$-factor corresponds to the case when $\kappa_{23}$ coincides with $\kappa_{12}$, as shown in \figref{fig:impact-of-kappa}. Similarly, when $\kappa_{23}$ is fixed, the sharpest curve occurs when $\kappa_{12} = \kappa_{23}$ as $\kappa_{12}$ is changed. Therefore, the sharpest curve is found when $\kappa_{12} = \kappa_{23} = \kappa$ in the experiment. % 此处应该补一个图

\begin{widetext}
% \onecolumngrid
\begin{figure}[!ht]
    \centering
    \includegraphics[width=7.0in]{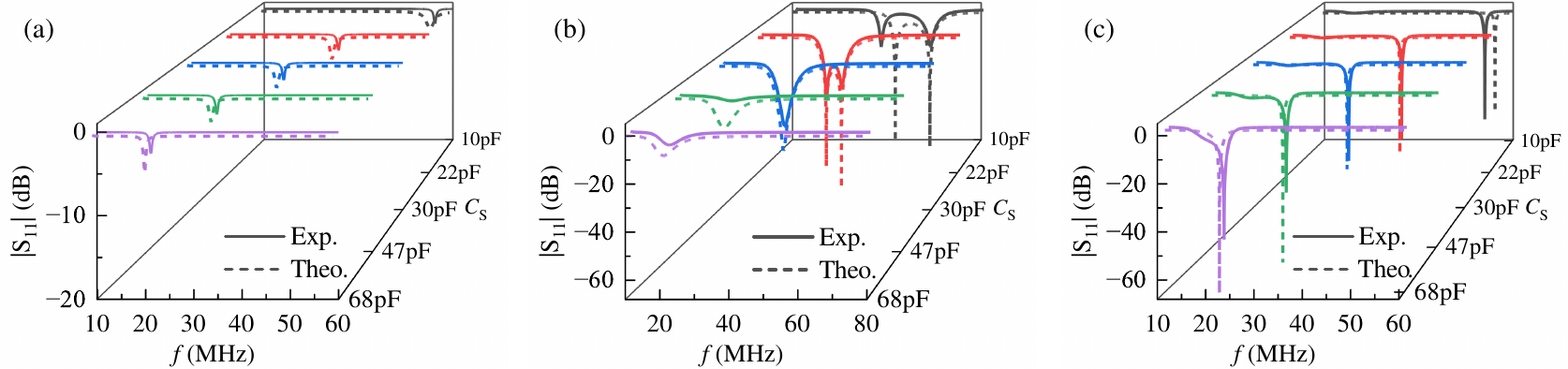}
    \makeatletter
    \renewcommand{\@makecaption}[2]{%
        \par\vskip\abovecaptionskip\begingroup\small\rmfamily
        \splittopskip=0pt
        \setbox\@tempboxa=\vbox{
            \@arrayparboxrestore \let \\ \@normalcr
            \hsize=2.1\hsize \advance\hsize-1em %\hsize=7.0in
            \let \\ \heading@cr
            \@make@capt@title {#1}{#2}}%
        \vbadness=10000
        \setbox\z@=\vsplit\@tempboxa to .55\ht\@tempboxa
        \setbox\z@=\vtop{\hrule height 0pt \unvbox\z@}
        \setbox\tw@=\vtop{\hrule height 0pt \unvbox\@tempboxa}
        \noindent\box\z@\hfill\box\tw@\par
        \endgroup\vskip \belowcaptionskip }
    \makeatother
    \caption[]{(color online) Comparison of the reflection spectra among (a) conventional, (b) second-order PT-symmetric and (c) the proposed sensing mechanism based on perturbed PT-symmetric electronic trimer when $\kappa_\text{trimer} = 0.2$ and $\kappa_\text{conv} = \kappa_\text{dimer} = 0.2\sqrt{2}$. Dashed and solid curves correspond to the theoretical and experimental results, respectively.}
    \label{fig:comparison_conventional_pt}
\end{figure}
% \twocolumngrid
\end{widetext}

We also compare the performance of the proposed wireless readout mechanism with the conventional method where only a readout antenna and an inductively coupled LC microsensor are incorporated \cite{chen2010wireless}, as well as with the standard second-order PT-symmetric readout system consisting of inductively coupled RLC resonators with balanced gain and loss respectively \cite{chen2018generalized}, as shown in \figref{fig:comparison_conventional_pt}. Note that we set the coupling coefficient between adjacent inductor coils of PT-symmetric trimer to be $\kappa_\text{trimer} = 0.2$ while keeping the same interrogation distance for conventional and second-order PT-symmetric systems which yields the corresponding coupling coefficient to be $\kappa_\text{conv} = \kappa_\text{dimer} = 0.2\sqrt{2}$, as explained in \appref{sec:append_c}. For the conventional readout method, the measured minimum of the reflection spectrum $|S_{11}|_{f = f_\text{dip}}$ is only greater than about $-5~\si{dB}$ [see \figref{fig:comparison_conventional_pt}(a)]. However, for the proposed perturbed third-order PT-symmetric wireless readout mechanism, $|S_{11}|_{f = f_\text{dip}}$ can reach $-50~\si{dB}$ which is ten times larger than that of conventional method, as shown in \figref{fig:comparison_conventional_pt}(c). The proposed readout scheme exhibits greater $Q$-factor compared to conventional scheme. Therefore, the proposed readout method enables finer spectral resolution in contrast with the conventional method under the same interrogation distance. Compared to standard second-order PT-symmetric wireless sensing systems, even though the sensitivity of the proposed third-order PT-symmetric system is relatively smaller than the second-order PT-symmetric system working in PT-symmetric phase nearby the exceptional point (EP), the readout capacitance range is larger. For the second-order PT-symmetric system, when the gain/loss parameter $\gamma = R\sqrt{C/L}$ is less than $\gamma_\text{EP} = [2 - 2(1-\kappa_\text{dimer} ^2)^{1/2}]^{1/2} = 0.286$ resulting in $C_s < C_\text{EP} = L (\gamma_\text{EP} / R)^2 =  29.4~\si{pF}$, the system is in PT-symmetric phase and the readout refection spectrum exhibits greater $Q$-factor, as can be seen from the black and red curves in \figref{fig:comparison_conventional_pt}(b). Especially, when $\gamma$ is close to the exceptional point $\gamma_\text{EP}$, the sensitivity can be much higher. Nevertheless, when $\gamma > \gamma_\text{EP}$ (\emph{i.e.}, $C > C_\text{EP}$), the symmetry is spontaneous breaking and the $Q$-factor will degrade significantly as $\gamma$ is increased, which can be seen from the green and purple curves in \figref{fig:comparison_conventional_pt}(b). Note that in the PT-broken phase, the reflection spectrum with high $Q$-factor [see blue curves in \figref{fig:comparison_conventional_pt}(b)] only occurs nearby EP when $\gamma$ is slightly larger than $\gamma_\text{EP}$, which has been reported in Ref.~\onlinecite{zhou2020enhancing}. Therefore, the readout capacitance range with great $Q$-factor for the second-order PT-symmetric system is in the range of $(0, C_\text{EP})$. However, for the proposed third-order PT-symmetric system, the readout capacitance range is $(0, C_\text{max})$ where $C_\text{max}$ can be set arbitrarily according to the specific application scenario to cover the whole quantity to be measured. In our experiment, the capacitance of gain/loss resonators is chosen as $C = 100~\si{pF}$, therefore the readout capacitance range is $(0, 100~\si{pF})$, while for second-order PT-based system the corresponding range will be $(0, 29.4~\si{pF})$.

\begin{figure}[!ht]
    \centering
    \includegraphics[width=3.4in]{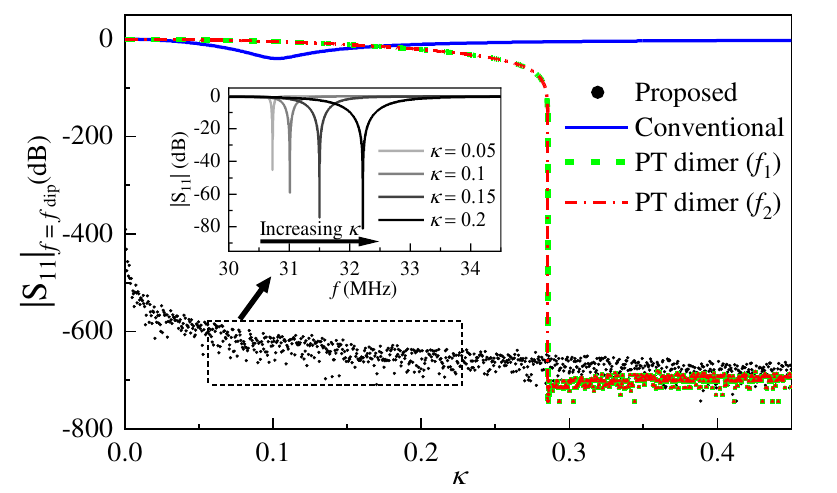}
    \caption{(color online) Minimum of $|S_{11}|$ comparison between conventional, standard PT-symmetric dimer and the proposed perturbed PT-symmetric electronic trimer when $C_s = 30~\si{pF}$ (In the standard PT-symmetric dimer, $C_r = C_s = 30~\si{pF}$, which yields gain/loss parameter $\gamma = 0.289$). The inset illustrates the reflection spectrum of the proposed perturbed third-order PT-symmetric readout mechanism with different coupling coefficient.}
    \label{fig:comparison_s11}
\end{figure}

Figure \ref{fig:comparison_s11} compares the theoretical reflection coefficient minimums $S_{11}^\text{min} = |S_{11}|_{f = f_\text{dip}}$ of the proposed perturbed third-order PT-symmetric wireless readout mechanism with the conventional and the standard second-order PT-symmetric readout system by substituting the analytical eigenfrequencies of each case into the reflection coefficient [see \eqref{eq:S11} in \appref{sec:append_b}]. For the conventional second-order PT-symmetric case, two eigenfrequencies are considered respectively, which are two real numbers in the PT-symmetric phase when $\kappa > \kappa_\text{EP}^\text{dimer}$ and a pair of complex conjugate numbers in the PT-broken phase when $\kappa < \kappa_\text{EP}^\text{dimer}$. Here, the range of the coupling coefficient $\kappa$ is $(0,0.45)$ for comparison in weak coupling regime; the sensor capacitance is fixed to $C_s = 30~\si{pF}$; and the sensor resistance of conventional system is $R = 0.5~\si{\ohm}$. Note that for standard second-order PT-based systems, it requires the reader capacitance $C_r = C_s = 30~\si{pF}$ and sensor resistance $R = Z_0 = 50~\si{\ohm}$ yielding the corresponding gain/loss parameter $\gamma = 0.289$. Comparing the blue solid lines and discrete black data points in \figref{fig:comparison_s11}, it is evident that the proposed perturbed third-order PT-symmetric readout mechanism exhibits much greater $|S_{11}|_{f = f_\text{dip}}$ than conventional method where high $Q$-factor only exists in a limited range of $\kappa$. Compared to the standard second-order PT-based system, the $|S_{11}|_{f = f_\text{dip}}$ of the proposed system can maintain large absolute value (high $Q$-factor) in the whole range of $\kappa$. However, for the conventional second-order PT-based system, high $Q$-factor only exists in the PT-symmetric phase when $\kappa > \kappa_\text{EP}^\text{dimer} = \gamma \sqrt{1-\gamma^2/4} = 0.286$ and in the PT-broken phase nearby EP \cite{zhou2020enhancing}, but will degrade significantly as $\kappa$ becomes small. Therefore, the proposed perturbed third-order PT-symmetric readout mechanism shows better performance in the weak coupling regime compared to the standard second-order PT-symmetric system. Furthermore, the minimum $S_{11}$ of the proposed system is decreasing as the coupling coefficient $\kappa$ is increased, implying deeper spectrum dip, which is originated from the stronger reflected waves from neutral and loss resonators. Note that even though the minimum $S_{11}$ can reach about $-700~\si{dB}$ for the proposed third-order and the standard second-order PT-based system in PT-symmetric phase in theory, practical measured spectrum can only reach several tens of decibels due to the limited sweep point which may skip the lowest dip. 

The inset of \figref{fig:comparison_s11} shows the theoretical reflection spectra evolution of the proposed readout mechanism as $\kappa$ increases from 0.05 to 0.2. It can be seen that both the dip frequency $f_\text{dip}$ and $Q$-factor (resonance line width of the dip) are dependent on the coupling coefficient $\kappa$. Specifically, the minimum of $|S_{11}|$ becomes smaller and the resonance line width becomes broader as $\kappa$ is increased. Moreover, the dip frequency $f_\text{dip}$ is approaching to the natural resonance frequency of the senor resonator, \emph{i.e.}, $f_s = 30.63~\si{MHz}$, as $\kappa$ is decreased. It should be pointed out that even though the position of the $|S_{11}|$ minimum is higher in the weak coupling regime, the line width is narrower implying higher spectral resolution, which validates the feasibility of the proposed PT-symmetric-trimer-based sensing mechanism in application scenarios where longer interrogation distances are required.

\section{Conclusion}\label{sec:concl}
In summary, we report a sandwich-type wireless capacitance readout system based on parity-time-symmetric electronic trimer consisting of gain-neutral-loss LC resonators chain. Theoretical analysis and experimental verification show that perturbing the neutral resonator of the trimer can realize wireless capacitance readout with high $Q$-factor thanks to the existence of a single real mode of the system in the weak coupling regime. Compared to conventional wireless sensing systems, the proposed mechanism exhibits finer spectral resolution due to the higher $Q$-factor. Different from the standard second-order \cite{chen2018generalized} and third-order \cite{sakhdari2019experimental} PT-symmetric systems, the proposed readout mechanism functions in the weak coupling regime, extending the interrogation distance between sensor and reader tank while maintaining ultrahigh resolution. Furthermore, the capacitors on gain and loss side do not need to be manually tuned to maintain the PT symmetry condition and the readout capacitance range is wider whose upper bound is determined by the capacitance of gain/loss LC tank. Thus, the proposed wireless sensing method is more suitable for application scenarios such as implanted medical devices where longer interrogation distance is usually required, providing a practical pathway toward the PT-enhanced wireless sensing.

\begin{acknowledgments}
T.D. acknowledges support from National Natural Science Foundation of China (NSFC) under grant no. 51977165.
\end{acknowledgments}

\appendix
\renewcommand{\thefigure}{\thesection\arabic{figure}}
\section{Derivation of System Equations for PT-symmetric Electronic Trimer}\label{sec:append_a}
\setcounter{figure}{0}  
Applying Kirchhoff's laws to the PT-symmetric trimer circuit shown in \figref{fig:schematics}(b), one can obtain the following systems equations as
\begin{subequations} \label{eq:kcl_kvl}
\begin{align} 
    -I_1 Z_0 + V_{C_1} + V_{L1} &=0, \\
    V_{C_2} + V_{L_2} &= 0, \\
    I_3 R + V_{C_3} + V_{L3} &= 0, 
\end{align}
\end{subequations}
where $V_{C_n}$ for $n = 1, 2, 3$ represent the capacitor voltage of series gain, neutral and loss resonators respectively; the $i-v$ relation of the inductors yields $V_{L1} = L I_1'(t) + M I_2'(t)$, $V_{L2} = L I_2'(t) + M [I_1'(t)+I_3'(t)]$ and $V_{L3} = L I_3'(t) + M I_2'(t)$; $I_n(t) = C_n V_{C_n}'(t)$ stands for   each inductor current; $I_n^\prime(t)$ [or $V_{C_n}'(t)$] denotes the first-order derivatives of $I_n(t)$ [or $V_{C_n}(t)$] with respect to the variable $t$. By defining the coupling coefficient $\kappa = L/M$, gain/loss parameter $\gamma = R \sqrt{C/L} = Z_0 \sqrt{C/L}$, natural resonance frequency $\omega_0 = 1/\sqrt{LC}$ and the normalized time (phase) $\tau = \omega_0 t$, the second order ordinary differential system equations can be derived as
\begin{widetext}
\begin{subequations} \label{eq:system_equation_differential}
  \begin{align} 
    I_1''(\tau) & = -\frac{1-\kappa^2}{1-2\kappa^2} I_1 + \frac{1}{\eta} \frac{\kappa}{1-2\kappa^2} I_2 - \frac{\kappa^2}{1-2\kappa^2} I_3 + \gamma \frac{1-\kappa^2}{1-2\kappa^2} I_1'(\tau) - \gamma \frac{\kappa^2}{1-2\kappa^2} I_3'(\tau),       \\
    I_2''(\tau) & = - \frac{1}{\eta (1-2\kappa^2)} I_2 + \frac{\kappa}{1-2\kappa^2} (I_1 + I_3) - \gamma \frac{\kappa}{1-2\kappa^2} I_1'(\tau) + \gamma \frac{\kappa}{1-2\kappa^2} I_3'(\tau),   \\
    I_3''(\tau) & = -\frac{1-\kappa^2}{1-2\kappa^2} I_3 + \frac{1}{\eta} \frac{\kappa}{1-2\kappa^2} I_2 - \frac{\kappa^2}{1-2\kappa^2} I_1 + \gamma \frac{\kappa^2}{1-2\kappa^2} I_1'(\tau) - \gamma \frac{1-\kappa^2}{1-2\kappa^2} I_3'(\tau), 
  \end{align}
\end{subequations}
\end{widetext}
which are three coupled second-order differential equations. Here, the primes ``$\prime$" and double primes ``$\prime\prime$" denotes the first- and second-order derivatives with respect to the normalized time $\tau$, respectively. By considering an $\exp(\text{i} \omega \tau)$ time dependence throughout and defining the state vector as $\Phi = (I_1, I_2, I_3, I'_1, I'_2, I'_3)^T$, the system equations can be recast into the Liouvillian formalism as given by \eqref{eq:sys-eq} and \eqref{eq:Liouvillian} in the main text.

\section{Derivation of the Reflection Coefficients for Conventional LC Sensors, Standard PT-symmetric Dimers and the Proposed Perturbed PT-symmetric Trimers} \label{sec:append_b}
\setcounter{figure}{0}  
For the sinusoidal steady-state circuit of the proposed wireless capacitance readout system shown in \figref{fig:schematics}(b), according to Kirchhoff's law, one can obtain
\begin{subequations} \label{eq:kirchhoff}
\begin{align}
    \text{i} (\omega - \omega^{-1}) I_3 + \gamma I_3 + \text{i} \omega \kappa I_2 &= 0,  \\
    \text{i} (\omega - \eta^{-1} \omega^{-1}) I_2 + \text{i} \omega \kappa (I_1 + I_3) &= 0,   \\
    \text{i} (\omega - \omega^{-1}) I_1 + \text{i} \omega \kappa I_2 & = \gamma V_\text{in}/R.
\end{align}
\end{subequations}
Here, $\omega$ is normalized by the natural resonance frequency $\omega_0 = 1/\sqrt{L C}$; $\gamma = R \sqrt{C/L}$ is the normalized gain/loss parameter; $\kappa = M/L$ is the coupling coefficient; and $\eta = C_s/C$ is the capacitive perturbation parameter. By evaluating for $Z_\text{in} = V_\text{in}/I_1$, the input impedance of the perturbed PT-symmetric trimer can be derived as
\begin{equation} \label{eq:input_impedance}
    Z_\text{in}^\text{trimer} = R \frac{ (1-\omega ^2) [h(\omega) - \eta \kappa^2 \omega^4] + \text{i} \gamma \omega h(\omega)}{\gamma \omega [\gamma \omega (\eta \omega^2-1) + \text{i} h(\omega)]},
\end{equation}
where $h(\omega) = \eta(1 - \kappa^2)\omega^4 - (1 + \eta)\omega^2 + 1$. Consequently, the reflection coefficient can then be calculated as 
\begin{equation} \label{eq:S11}
    S_{11} = 20 \lg \left|\frac{Z_\text{in}-Z_0}{Z_\text{in}+Z_0}\right|, %_{Z_0=50~\si{\ohm}}, 
\end{equation}
where $Z_0=50~\si{\ohm}$ is the intrinsic impedance of the VNA.

\begin{figure}[!ht]
    \centering
    \includegraphics[width=3.4in]{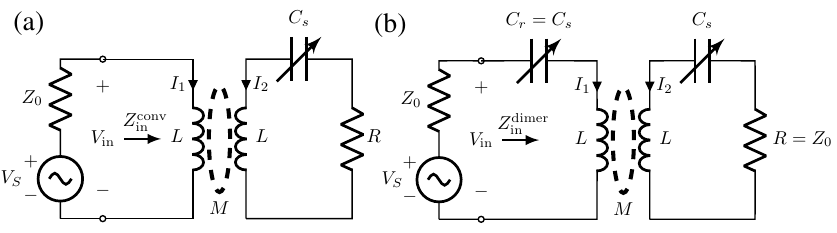}
    \caption{(a) Conventional LC wireless sensor readout mechanism and (b) PT-symmetric LC wireless sensor with second-order EP.}
    \label{fig:conv_dimer}
\end{figure}

As a reference, \figref{fig:conv_dimer}(a) illustrates the schematic circuit for a conventional LC wireless readout system, where the sensor coil together with a capacitive sensor is inductively coupled to the readout coil.
The value of the capacitive sensor is measured through the dip shift of the reflection spectrum. Now, the Kirchhoff's law yields  $\text{i}(\omega - \omega^{-1})I_2 + \gamma I_2 + \text{i} \omega \kappa I_1 = 0$ and $V_\text{in} \gamma/R = \text{i} \omega I_1  + \text{i} \omega \kappa I_2$; thus, the input impedance $Z_\text{in} = V_\text{in}/I_1$ can be expressed as
\begin{equation} \label{eq:input_imp}
    Z_\text{in}^\text{conv} = R \frac{\omega [\text{i} \gamma  \omega -(1-\kappa^2) \omega^2+1]}{\gamma [\gamma \omega -\text{i} (1-\omega^2)]}.
\end{equation}
For the PT-symmetric sensor with second order EP, whose schematic circuit is shown in \figref{fig:conv_dimer}(b), the reader circuit is replaced by a RLC series resonator with negative resistance. The Kirchhoff's law yields $\text{i}(\omega - \omega^{-1})I_2 + \gamma I_2 + \text{i} \omega \kappa I_1 = 0$ and $V_\text{in} \gamma/R = \text{i}(\omega-\omega^{-1}) I_1 + \text{i} \omega \kappa I_2$; and the input impedance $Z_\text{in}=V_\text{in}/I_1$ now reads as
\begin{equation} \label{eq:input_imp1}
    Z_\text{in}^\text{dimer} = R \frac{\text{i} \gamma  (\omega^2-1) \omega - (1-\kappa^2) \omega^4 + 2 \omega^2-1}{\gamma \omega  [\gamma \omega - \text{i} (1-\omega^2)]}.
\end{equation}
When the values of $R$, $Z_0$, $L$, $\gamma$ and $\kappa$ are known, the theoretical frequency response of the reflection coefficient $S_{11}$ defined in \eqref{eq:S11} can be obtained, as demonstrated in \figref{fig:comparison_conventional_pt} of the main text.

\section{Coils Deign, Relation Between the Coupling Coefficients of Concentrically Aligned Coil Dimers and Trimers} \label{sec:append_c}
\setcounter{figure}{0}
In order to validate the proposed readout mechanism in radio-frequency range, we designed a planar spiral inductor on board, as illustrated in \figref{fig:coupling}(a). The self-inductance of printed spiral coils is obtained from the ratio of the magnetic flux generated by the conductor to the current, which can be approximated by the closed form as \cite{mohan1999simple}
\begin{equation} \label{eq:L_calculation}
    L \approx 0.5 \mu_\text{0} N^2 \phi_\text{avg} \left( 0.9 + 0.2 f^2 - \ln f \right),
\end{equation}
where $\mu_\text{0}$ is the vacuum permeability; $N$ is the number of turns; $\phi_\text{avg} = 2 r_\text{in} + N(w+s)$ is the average diameter of the coil, $r_\text{in}$ denoting inner radius, $w$ and $s$ being the width and spacing of the spiral coil, respectively; $f = (r_\text{out} + r_\text{in})/(r_\text{out} - r_\text{in})$ is filling rate with $r_\text{out}$ being the outer radius of the spiral coil.
\begin{figure}[!ht]
    \centering
    \includegraphics[width=3.4in]{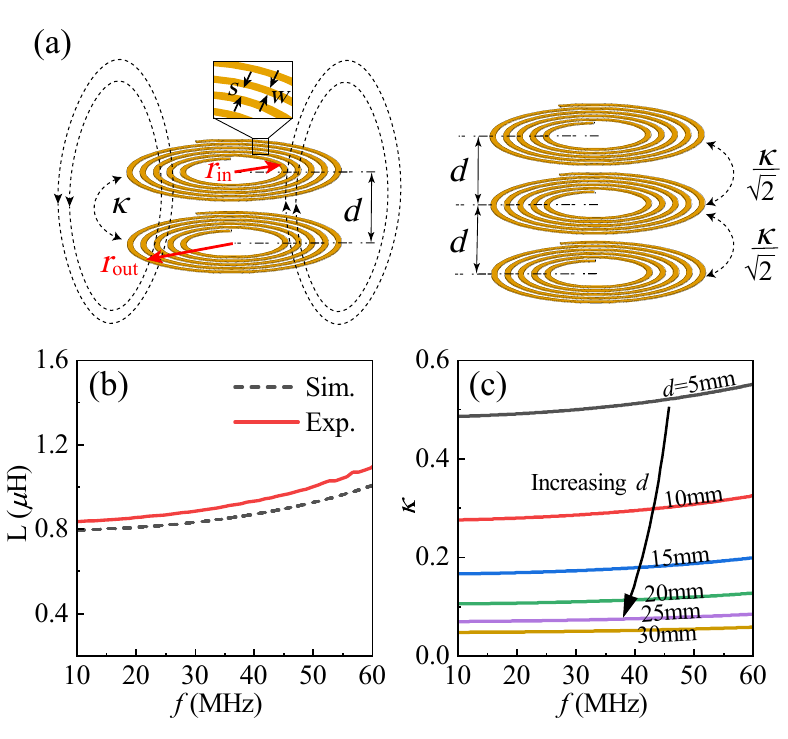}
    \caption{(color online) (a) Illustration of concentrically aligned printed spiral coil dimer (left) and trimer (right). When the separations distance of two adjacent coils are equal, the coupling coefficient of the trimer is that of dimer divided by $\sqrt{2}$. (b) Measured (solid curves) and simulated (dashed curves) self-inductance of the designed coils. (c) The coupling coefficients under different working frequencies for various separation distance.}
    \label{fig:coupling}
\end{figure}

Due to the existence of parasitic capacitance affected by coil size, the inductance would increase with frequency. To realize the frequency-stable and almost non-dispersive target inductance of $L = 0.9~\si{\micro\henry}$ at a frequency range from $10~\si{MHz}$ to $60~\si{MHz}$, the geometric parameters of the coils reads as: $N = 5.05$, $r_\text{in} = 8~\si{mm}$ and $s= w = 1.188~\si{mm}$. Figure \ref{fig:coupling}(b) shows the simulated and measured inductance of the designed coils as a function of frequency, which further validates our design scheme. We also carried out full-wave simulation in order to evaluate the mutual inductance $M$ between two coaxial coils, and further obtain the coupling coefficient $\kappa = M / L$, as illustrated in \figref{fig:coupling}(c). It can be seen that the coupling coefficient is decreased as the distance between coaxial coils is increased. In the weak coupling region when the separation distance $d$ is large, the coupling coefficient is almost frequency-stable.

It should also be pointed out that for the same interrogation distance, the coupling coefficient between two concentric coils and adjacent coupling of three concentric coils is different. Assuming the inductances of the three coils in \figref{fig:coupling}(a) are $L_1 = L_2 = L_3 = L$ (``1" for the gain, ``2" for the neutral LC tank and ``3" for the loss), and the mutual inductances between adjacent coils are $M_\text{12} = M_{23} = M$. Thus, the energy stored in the inductors reads $W = L(I_1^2 + I_2^2 + I_3^2)/2 + M I_1 I_2 + M I_2 I_3$, which can be expressed in the matrix form as
\begin{equation} \label{eq:E_calculation}
    W = \frac{1}{2} I^{\mathrm{T}} \mathcal{M} I,
\end{equation}
where
\begin{equation} \label{eq:mutual-inductance-matrix}
    \mathcal{M} = \begin{pmatrix}
            L  &  M  &  0 \\
            M  &  L  &  M \\
            0  &  M  &  L
        \end{pmatrix}
\end{equation}
is the mutual inductance matrix and $I = (I_1, I_2, I_3)^\text{T}$. Due to the arbitrariness of the current, the inductance matrix must be positive definite to guarantee that the energy is positive, which yields $L^3 - 2 L M^2>0$. Therefore, for the concentrically aligned coil trimer, $\kappa = M/L < \sqrt{2} / 2$. As a result, if the coupling coefficient reads $\kappa$ for the coil dimer when the separation distance is $d$, the coupling coefficient becomes $\kappa / \sqrt{2}$ for the concentric coil trimer with the same adjacent separation distances.

\bibliography{main}

\end{document}